\documentclass[sn-apa,Numbered]{sn-jnl}

\usepackage{graphicx}%
\usepackage{multirow}%
\usepackage{amsmath,amssymb,amsfonts}%
\usepackage{amsthm}%
\usepackage{mathrsfs}%
\usepackage[title]{appendix}%
\usepackage{xcolor}%
\usepackage{textcomp}%
\usepackage{manyfoot}%
\usepackage{booktabs}
\usepackage{array} 
\usepackage{algorithm}%
\usepackage{algorithmicx}%
\usepackage{algpseudocode}%
\usepackage{float} 

\usepackage{csquotes} 
\usepackage{comment}

\usepackage{listings}  
\definecolor{codegreen}{rgb}{0,0.6,0}
\definecolor{codegray}{rgb}{0.5,0.5,0.5}
\definecolor{codepurple}{rgb}{0.58,0,0.82}
\definecolor{backcolour}{rgb}{0.95,0.95,0.92}
\definecolor{royalblue}{RGB}{65,105,225}
\lstdefinestyle{mystyle}{
    backgroundcolor=\color{backcolour},   
    commentstyle=\color{codegreen},
    keywordstyle=\color{royalblue},
    numberstyle=\tiny\color{codegray},
    stringstyle=\color{codepurple},
    basicstyle=\ttfamily\footnotesize,
    breakatwhitespace=false,         
    breaklines=true,                 
    captionpos=b,                    
    keepspaces=true,                 
    numbers=left,                    
    numbersep=5pt,                  
    showspaces=false,                
    showstringspaces=false,
    showtabs=false,                  
    tabsize=2
}
\theoremstyle{thmstyleone}%
\theoremstyle{thmstyletwo}%

\theoremstyle{thmstylethree}%

\usepackage{colortbl}
\definecolor{lightgray}{RGB}{224,224,224}
\definecolor{lightcyan}{RGB}{201,237,255}

\begin{document}

\title[Article Title]{Examining Differential Item Functioning in Self-Reported Health Survey Data: Via Multilevel Modeling}


\author*[1]{\fnm{Dandan} \sur{Chen Kaptur}}\email{danielle.chen@pearson.com}
\author[2]{\fnm{Yiqing} \sur{Liu}}
\author[3]{\fnm{Bradley} \sur{Kaptur}}
\author[4]{\fnm{Nicholas} \sur{Peterman}}

\author[5]{\fnm{Jinming} \sur{Zhang}}
\author[5]{\fnm{Justin} \sur{Kern}}
\author[5]{\fnm{Carolyn} \sur{Anderson}}

\affil*[1]{\orgname{Pearson}. ORCID: 0000-0002-4020-3422}
\affil[2]{\orgname{Stanford University}}
\affil[3]{\orgname{HSHS Saint John's Hospital}. ORCID: 0000-0003-4142-0231}
\affil[4]{\orgname{Carle Illinois College of Medicine}}
\affil[5]{\orgname{University of Illinois Urbana-Champaign}}


\abstract{Few health-related constructs or measures have received a critical evaluation in terms of measurement equivalence, such as self-reported health survey data. Differential item functioning (DIF) analysis is crucial for evaluating measurement equivalence in self-reported health surveys, which are often hierarchical in structure. Traditional single-level DIF methods in this case fall short, making multilevel models a better alternative. We highlight the benefits of multilevel modeling for DIF analysis, when applying a health survey data set to multilevel binary logistic regression (for analyzing binary response data) and multilevel multinominal logistic regression (for analyzing polytomous response data), and comparing them with their single-level counterparts. Our findings show that multilevel models fit better and explain more variance than single-level models. This article is expected to raise awareness of multilevel modeling and help healthcare researchers and practitioners understand the use of multilevel modeling for DIF analysis.}




This is a preprint. The Version of Record of this article is published in \textit{Quality of Life Research}, and
is available online at \url{https://link.springer.com/article/10.1007/s11136-025-03936-9}.

\maketitle

\section{Introduction}

To examine the disproportionate impact of health interventions or other determinants on health outcomes for certain populations, including disadvantaged minorities, it is crucial to establish measurement equivalence when evaluating outcomes across groups differing in gender, ethnicity, and other demographic factors. While validating psychometric instruments has been a routine practice in research, critical evaluations of measurement equivalence, particularly through differential item functioning (DIF) analysis, have historically been limited \citep{teresi_differential_2007}. In recent years, however, there has been significant progress in this area, with a growing body of research employing DIF analysis to investigate measurement equivalence \citep[e.g.,][]{rice_preliminary_2020,rouquette_validity_2018,quistberg_building_2019,tiego_precision_2023,jones_differential_2019,teresiDifferentialItemFunctioning2021}. 

This growing research underscores the need for continued evaluation of health constructs and measures to ensure fairness and validity in assessing health outcomes for diverse populations. It is worth noting that some health studies did not clearly describe the method used for DIF analysis \citep[e.g.,][]{rouquette_validity_2018}, and conventional DIF methods based on single-level models were frequently used to analyze health survey data with a nesting structure \citep[e.g.,][]{rice_preliminary_2020}, for which recent DIF methods based on multilevel models are better alternatives in terms of accuracy of DIF estimation. By employing advanced methodologies for DIF, researchers can better evaluate health measures to ensure fairness and validity in assessing health outcomes for diverse populations. 

\section{Challenge: Examine DIF in self-reported health survey data}

Self-reported health survey data encapsulate rich information about physical, mental, and social health trends of the studied population \citep[e.g.,][]{loremSelfreportedHealthPredictor2020,wuorelaSelfratedHealthObjective2020}. However, its inherent subjectivity makes it susceptible to geographical, cultural, and socioeconomic biases, which can threaten the validity of the data and pose challenges in getting to know the truth \citep[e.g.,][]{ulitzschDifferencesResponsescaleUsage2024}. The utilization of self-reported health survey data in public health research and interventions necessitates a rigorous examination of the validity of this data. It underscores the importance of adopting appropriate methods that can navigate the complexities associated with self-reported health data and improve the accuracy of our findings.

Existing research has shown that bias can manifest itself as construct bias, method bias, and item bias \citep[e.g.,][]{van_de_vijver_towards_1997,werner_translating_1970}. The framework of differential item functioning \citep[DIF;][]{berk_handbook_1982} underpins many modern methods to investigate item bias, specifically. In the psychometric literature, DIF analysis examines whether there is a systematic between-group difference in examinees' probability to answer an item correctly when controlling for their ability \citep{american_educational_research_association_standards_2014}. 
While DIF is not equivalent to item bias, the former can provide evidence for the latter when it arises from content or characteristics of the item that are unrelated to the intended construct being measured \citep{penfield_item_2006,lee_effects_2017}. Some existing DIF methods examine both uniform and nonuniform DIF while many others do not. \textit{Uniform DIF} occurs when respondents from one group consistently outperform respondents of the same latent trait from another group; \textit{nonuniform DIF} exists when this performance difference is not consistent \citep{mellenbergh_contingency_1982}. 

To examine systematic between-group differences in self-reported health survey data, we can apply existing DIF methods by generalizing DIF-related concepts via the following modifications:
\begin{enumerate}
    \item focusing on the response of interest instead of a \enquote{correct} response,
    \item describing the subjects who provide responses as \enquote{respondents} instead of \enquote{examinees,}
    \item using a latent trait variable instead of \enquote{ability,} selecting one closely related to the probability of obtaining the response of interest (similar to how \enquote{ability} relates to the probability of a correct response in the conventional DIF framework).
\end{enumerate}
The corresponding terminology may shift in the applications of DIF to address the unique contexts in health research. 

Most DIF methods currently used in health research, as well as in the broader DIF literature, are based on single-level modeling \citep[e.g.,][]{rice_preliminary_2020}. They are not best suited for self-reported health survey data that has a nesting or hierarchical structure (e.g., individual responses are nested within larger units differentiated by population density, economy, and culture). These single-level DIF methods include the logistic regression procedure \citep[LR;][]{swaminathan_detecting_1990}, the Mantel-Haenszel procedure \citep[MH;][]{holland_alternate_1985}, Lord’s Wald test \citep{lord_study_1976}, and the IRT likelihood-ratio test \citep[IRT-LR;][]{thissen_beyond_1986}. The first two do not require the IRT model and are labeled with \enquote{the observed-score approach,} whereas the latter two require the IRT model and are categorized as \enquote{the IRT approach} \citep{osterlind_differential_2009,lee_lords_2015}. An extensive review of these methods can be found in \citet{chenModelingItemBias2023a}. 

Single-level modeling approaches have at least three problems when the response data has a hierarchical structure. First, they waste information and render inaccurate estimates because of the lack of consideration of variances between hierarchies of the data \citep{raudenbush_hierarchical_2002}. Second, they do not help model the effect of higher-level variables associated with DIF on the outcome variable at the lower level \citep{gelman_data_2007}. Third, their model estimates are limited to describing the sample and are not generalizable to describe the population \citep{shadish_experimental_2002}. Moreover, most of the single-level DIF methods are limited to the analysis of two groups at a time, showing inflated error rates for DIF analyses involving more than two groups \citep{chenModelingItemBias2023a,woods_langer-improved_2013,penfield_assessing_2001}.

\section{Solution: Multilevel modeling for DIF analysis}

For the self-reported health survey data that is hierarchical in nature, multilevel modeling has been demonstrated to be a superior alternative to single-level modeling. It effectively addresses issues that arise in DIF methods built on single-level models. First, multilevel modeling is advantageous when fundamental assumptions in single-level modeling such as independence and homoscedasticity are violated, as is often the case with hierarchical data \citep{raudenbush_hierarchical_2002}. In this regard, multilevel modeling can account for correlated errors resulting from the dependency of individual observations within a nested structure, and it does not require the same sample size or variance for higher-level units due to its randomization approach \citep{snijders_multilevel_2011}. 

Second, multilevel modeling allows for the assessment of higher-level variables' impact on DIF \citep{gelman_data_2007}. For instance, it can evaluate how cluster-level factors, like access to healthcare facilities, influence individual responses in health surveys, and identify whether these cluster characteristics contribute to DIF observed between population groups (e.g., gender, ethnicity, location) at the lower level of the model. Third, results from multilevel modeling are generalizable to the population because it incorporates random effects from higher-level clusters \citep{shadish_experimental_2002}. This ensures that conclusions about DIF in self-reported response data are applicable to a broader range of demographic groups. Fourth, multilevel modeling accommodates multi-group analyses by using multiple dummy-coded variables to cover more than two groups \citep{swanson_analysis_2002}. For instance, it can include multiple age groups in health data, allowing researchers to analyze and compare health outcomes across various age categories while accounting for cluster-specific random effects.

The first published work on DIF using multilevel modeling was by \citet{swanson_analysis_2002}, who adopted a multilevel binary LR model. This approach significantly improved the accuracy of DIF estimates and pooled information across items to explain sources of DIF, when including random item effects. Subsequently, \citet{den_noortgate_assessing_2005} proposed a series of multilevel models for DIF analysis, grounded in the Rasch model within item response theory (IRT). These models demonstrated the potential to integrate random effects associated with respondents, items, or other factors, along with item- or group-related covariates, to explain DIF. \citet{french_extensions_2013} proposed a multilevel MH statistic that adjusts the traditional MH statistic using a ratio of random variances from the multilevel model from \citet{swanson_analysis_2002}. \citet{huang_wald_2024} introduced a multilevel Wald test for polytomous items based on Lord’s Wald $\chi^2$ test.

Each of these multilevel approaches offers unique benefits, but they also come with specific limitations. For instance, the models from \citet{den_noortgate_assessing_2005} leverage the rigor of IRT, allowing for a deeper integration of item- and respondent-level characteristics and providing insights into hierarchical data structures. However, these models rely on assumptions such as local independence of item responses and monotonicity of response probabilities, which can be restrictive or challenging to meet in certain applications \citep{lord_applications_1980}. In contrast, the multilevel LR model adapted from \citet{swanson_analysis_2002} offers greater versatility by not requiring such stringent assumptions, making it suitable for contexts where IRT assumptions might not hold.

The choice of method depends on the research objectives and the nature of the data. For example, the IRT-based multilevel models proposed by \citet{den_noortgate_assessing_2005} are particularly useful when the goal is to investigate complex random effects structures or when there is a need for a model grounded in the theoretical framework of IRT. However, these models may be computationally intensive and less intuitive for practitioners unfamiliar with IRT. Conversely, the multilevel LR model provides a balance of flexibility and interpretability, making it a practical choice for broader applications. 

In the following sections, we demonstrate the use of two types of multilevel models corresponding to binary and polytomous response data, based on \citet{swanson_analysis_2002} because of the versatility. First, for analyzing DIF in \textit{binary} response data from a survey item, we utilize the \textit{multilevel binary LR} adapted from \citet{swanson_analysis_2002}. Its single-level counterpart, the LR procedure \citep{swaminathan_detecting_1990}, is reviewed in Appendix A, hereafter referred to as \enquote{traditional LR}. 
Second, for analyzing DIF in \textit{polytomous} response data, we propose the \textit{multilevel multinominal LR}. Its single-level counterpart is the multinominal LR, a type of multicategory logit models for unordered categories and is seen as a generalization of the binary LR \citep{agresti_introduction_2019}. Throughout this writing, we use the word \enquote{group} when referring to the group membership variable for group comparisons in DIF analysis, and the word \enquote{cluster} when referring to units of analyses across levels in the hierarchical data structure in multilevel modeling.


\subsection{Multilevel Binary LR for Binary Responses} 

The multilevel binary LR \citep{swanson_analysis_2002} was initially designed to include random effects associated with items, accounting for variations among items and offering more accurate DIF estimates compared to the traditional LR \citep{swaminathan_detecting_1990}. We have modified its notation to allow the incorporation of more than one focal group and of random effects from higher-level clusters in hierarchical data (e.g., access to healthcare, geographical locations, age groups, education levels). 

In this modified version, the level-1 submodel pertains to responses to a studied item from individual respondents, indexed by $p$, within a level-2 stratum, indexed by $i$. This two-level binary LR model is specified as follows:
\begin{equation}
\label{eqn:mlm_lr}
    \begin{array}{lrll}
        \textbf{Level 1:} & & &\\
         & \mbox{logit}(\pi_{pi}) &=& \mbox{log}(\dfrac{\pi_{pi}}{1-\pi_{pi}}) = \beta_{0i} + \beta_{1i} \theta_p + \boldsymbol{\beta_{2i} g_p}, \\
        \textbf{Level 2:} & & &\\
         & \beta_{0i} &=& \delta_{00}+U_{0i},  \\
         & \beta_{1i} &=& \delta_{10}+U_{1i},  \\
         & \boldsymbol{\beta_{2i}} &=& \boldsymbol{\delta_{20}} + U_{2i}.
    \end{array}
\end{equation}

\noindent where

\indent \hangindent=0.4in $\pi_{pi}$ is the probability for respondent $p$ to provide the targeted response to the studied item (i.e., the \enquote{probability targeted});

\indent \hangindent=0.4in $\beta_{0i}$ denotes the log-odds of the \enquote{probability targeted} in the reference group;

\indent \hangindent=0.4in $\beta_{1i}$ is the effect of respondents' latent trait on the log-odds of the \enquote{probability targeted} in the reference group;

\indent \hangindent=0.4in $\boldsymbol{\beta_{2i}}$ is a vector for the deviation effect in a focal group from the reference group;

\indent \hangindent=0.4in $\theta_p$ is the latent trait of respondent $p$ (e.g., it could be proxied by a continuous variable concerning the respondent's overall health status);

\indent \hangindent=0.4in $\boldsymbol{g_p}$ is a vector containing dummy-coded variables representing the group membership of respondent $p$, specified as $\begin{bmatrix}g_{p1} & g_{p2} & \cdots & g_{pJ}\end{bmatrix}$ ($g_{pj} = 1$ for being in the target group, indexed by $j = 1, 2, \cdots, J$);

\indent \hangindent=0.4in $\delta_{00}$ is the average log-odds of the \enquote{probability targeted;}

\indent \hangindent=0.4in $\delta_{10}$ is the average effect associated with respondents' latent trait;

\indent \hangindent=0.4in $\boldsymbol{\delta_{20}}$ is a vector for the average effect associated with being in each focal group;

\indent \hangindent=0.4in $U_{0i}$ is the random effect associated with level-2 clusters on the intercept $\beta_{0i}$;

\indent \hangindent=0.4in $U_{1i}$ and $U_{2i}$ denote the random effect associated with level-2 clusters on the slopes $\beta_{1i}$ and $\boldsymbol{\beta_{2i}}$, respectively;

\indent \hangindent=0.4in $U_{0i}$, $U_{1i}$, and $U_{2i}$ follow a multivariate normal distribution with the mean being 0.

The level-1 submodel of the multilevel binary LR is a binary LR model, similar to the one in traditional LR but without an ability-group interaction term. It is limited to binary response data analysis. The level-2 submodel in the multilevel binary LR accounts for random effects associated with higher-level clusters. When using this multilevel binary LR, the studied item exhibits DIF if any coefficient value in the vector $\boldsymbol{\beta_{2i}}$ is significantly different from 0.

\subsection{Multilevel Multinominal LR for Polytomous Responses} 

We propose the multilevel multinominal LR for analyzing DIF in polytomous response data. Its level-1 submodel is the multinominal extension of the traditional LR but without the ability-group interation term. Its level-2 submodel incorporates random effects tied to higher-level clusters. This two-level multilevel multinominal LR is written as

\begin{equation}
\label{eqn:mlm_poly}
    \begin{array}{lrll}
        \textbf{Level 1:} & & &\\
         & \mbox{log}(\dfrac{\pi_{c.pi}}{\pi_{B.pi}}) &=& \beta_{0i} + \beta_{1i} \theta_p + \boldsymbol{\beta_{2i} g_p}, \\
        \textbf{Level 2:} & & &\\
         & \beta_{0i} &=& \delta_{00}+U_{0i},  \\
         & \beta_{1i} &=& \delta_{10}+U_{1i},  \\
         & \boldsymbol{\beta_{2i}} &=& \boldsymbol{\delta_{20}} + U_{2i}.
    \end{array}
\end{equation}

\noindent where

\indent \hangindent=0.4in $\pi_{c.pi}$ is the probability for respondent $p$ to provide a non-baseline response, indexed by $c$, to the studied item ($c=1, 2, \cdots, C$);

\indent \hangindent=0.4in $\pi_{B.pi}$ is the probability for respondent $p$ to provide the baseline response, denoted by $B$, to the studied item.

\noindent The level-1 submodel pairs a selected baseline response with each of the other responses to the survey item, and it has $C$ equations, corresponding to the number of non-baseline responses. Parameters for these equations are estimated separately. For example, when the response data has three unique responses (e.g., 0, 1, and 2) and we choose one as the baseline (e.g., 0), $C=2$ and the level-1 submodel is written into
\[
\begin{array}{lrll}

\text{Equation 1:} & \mbox{log}(\dfrac{\pi_{1.pi}}{\pi_{B.pi}}) &=& \beta_{1.0i} + \beta_{1.1i} \theta_p + \boldsymbol{\beta_{1.2i} g_p}, \\
\text{Equation 2:}& \mbox{log}(\dfrac{\pi_{2.pi}}{\pi_{B.pi}}) &=& \beta_{2.0i} + \beta_{2.1i} \theta_p + \boldsymbol{\beta_{2.2i} g_p}. \\

\end{array}
\]

\section{Example: Analysis of population-density DIF in NSDUH data}

Existing research has demonstrated that population density can affect the quality of life and health perceptions \citep{douglasExaminingRelationshipUrban2022,zanderPerceivedHeatStress2018,fassioHealthQualityLife2013,waltonRelationshipsPopulationDensity2008}. We utilize the NSDUH data released in 2022 to examine population-density-related DIF in the self-reported health survey questions evaluating depression. This data provides comprehensive information on substance use and mental health trends among the U.S. population aged 12 and older. To show the impact of multilevel modeling on DIF results, below we compare the results from the multilevel binary LR and the multilevel multinominal LR with their single-level counterparts. We use respondents' education levels as the level-2 clusters in multilevel modeling, for existing research has unveiled that an additional year of education could reduce the likelihood of reporting depression and anxiety \citep{kondirolliMentalHealthEffects2022}. 

Specifically, we want to examine whether DIF exists for respondents from high and low population density areas in the probability to answer \enquote{Yes} to a selected NSDUH survey item. Hence, the targeted response in this DIF analysis is \enquote{Yes}. This item asks respondents the following question: \enquote{Have you ever in your life had a period of time lasting several days or longer when most of the day you felt sad, empty or depressed?} The latent trait variable to be controlled for is respondents' psychological distress level over the past 30 days, proxied by a score based on their responses to another six survey questions. We acknowledge the potential for DIF in this derived score; however, we do not examine DIF in this instance to maintain our focus on demonstrating the application of multilevel modeling. 
The reference group is the respondents from areas with low population density, and there are two focal groups: one is the respondents from areas with high population density, and the other is from non-classified areas.

\subsection{Data preparation}

The original NSDUH data file contains 2,605 variables and 59,069 respondents, composed by 27,047 males and 32,022 females aged 12 and older. We obtained two different data sets from this original data file. In the first data set, we reserve responses recoded with 1 for the response \enquote{Yes} and 0 for \enquote{No}, retain 45,827 eligible respondents to enable multilevel binary LR. In the second data set, we include responses recoded with 2 for the other valid responses including \enquote{Don't know}, \enquote{Refused}, and blank answers, retaining 47,094 eligible respondents for multilevel multinominal LR. Both the data sets contain no missing values. Table \ref{tab:ns} displays sample sizes by level-2 cluster.

\begin{table}
\centering
\caption{Sample size of respondents, by cluster.}
\label{tab:ns}
\begin{tabular}{>{\raggedleft\arraybackslash}p{0.12\linewidth} >{\raggedleft\arraybackslash}p{0.12\linewidth} >{\raggedleft\arraybackslash}p{0.12\linewidth} >{\raggedleft\arraybackslash}p{0.12\linewidth} >{\raggedleft\arraybackslash}p{0.07\linewidth} >{\raggedleft\arraybackslash}p{0.07\linewidth} >{\raggedleft\arraybackslash}p{0.08\linewidth}}  
    \toprule
    \textbf{Level-2 clusters} & \multicolumn{3}{c}{\textbf{Group}} & \multicolumn{3}{c}{\textbf{Response}} \\ 
     \cmidrule(lr){2-4} \cmidrule(lr){5-7}
    Edu. level & (0) Low-density area & (1) High-density area & (2) Non-classified area & (0) No & (1) Yes & (2) Others\\ 
    \midrule
    \multicolumn{7}{l}{\textbf{Data Set 1: With binary responses}}\\

\hspace{1em}1 & 126 & 140 & 10 & 232 & 44 & \\

\hspace{1em}2 & 75 & 115 & 5 & 177 & 18 & \\

\hspace{1em}3 & 45 & 49 & 7 & 82 & 19 & \\

\hspace{1em}4 & 158 & 102 & 56 & 240 & 76 & \\

\hspace{1em}5 & 314 & 206 & 39 & 414 & 145 & \\

\hspace{1em}6 & 407 & 273 & 57 & 529 & 208 & \\

\hspace{1em}7 & 1228 & 1059 & 142 & 1691 & 738 & \\

\hspace{1em}8 & 6488 & 4430 & 746 & 7942 & 3722 & \\

\hspace{1em}9 & 5208 & 3895 & 447 & 5607 & 3943 & \\

\hspace{1em}10 & 2286 & 1563 & 226 & 2616 & 1459 & \\

\hspace{1em}11 & 7630 & 7874 & 421 & 10029 & 5896 & \\
\multicolumn{7}{l}{}\\
\multicolumn{7}{l}{\textbf{Data Set 2: With polytomous responses}}\\

\hspace{1em}1 & 129 & 148 & 10 & 232 & 44 & 11\\

\hspace{1em}2 & 77 & 120 & 5 & 177 & 18 & 7\\

\hspace{1em}3 & 47 & 52 & 7 & 82 & 19 & 5\\

\hspace{1em}4 & 165 & 104 & 58 & 240 & 76 & 11\\

\hspace{1em}5 & 329 & 214 & 40 & 414 & 145 & 24\\

\hspace{1em}6 & 422 & 289 & 57 & 529 & 208 & 31\\

\hspace{1em}7 & 1270 & 1082 & 146 & 1691 & 738 & 69\\

\hspace{1em}8 & 6716 & 4579 & 769 & 7942 & 3722 & 400\\

\hspace{1em}9 & 5337 & 3985 & 456 & 5607 & 3943 & 228\\

\hspace{1em}10 & 2350 & 1603 & 232 & 2616 & 1459 & 110\\

\hspace{1em}11 & 7791 & 8075 & 430 & 10029 & 5896 & 371\\
    \bottomrule
    \end{tabular}
        \begin{tablenotes}  
            \item \hspace{-15pt} \textit{Note}. Numbers in parentheses in the header are codes in the data.
        \end{tablenotes}
\end{table}

We keep the following four variables for DIF analysis: 
\begin{enumerate}
    \item The item response variable (named \enquote{ADDPREV} in the survey). It contains responses 0 and 1 for multilevel binary LR modeling, and it include the response 2 for multilevel multinominal LR modeling.  
    \item The group membership variable (named \enquote{PDEN10}). It indicates the population density of the area where respondents are residing. It is recoded to have 0 for \enquote{low-density} (i.e., with fewer than one million people), 1 for \enquote{high-density} (i.e., with one million or more people), and 2 for areas that are not involved in Core Based Statistical Area (CBSA) classifications. In modeling, this variable is dummy-coded.
    \item The latent trait variable (named \enquote{KSSLR6MON}). It is a score that describes respondents' distress, ranging from 0 to 24. This score is closely associated with the probability of obtaining the response of interest in the item response variable.
    \item The level-2 variable (named \enquote{IREDUHIGHST2}). It describes respondents' education level, and there are 11 distinct levels, ranging from 1 for \enquote{completing fifth grade or less} to 11 for \enquote{completing college or more advanced education.}
\end{enumerate}
Appendix B displays information from the NSDUH codebook about how the selected categorical variables were originally coded. Appendix C.1 documents the R code for this data preparation. 

\subsection{Analysis and Results}

The modeling results are in Table \ref{tab:models1} and Table \ref{tab:models2}. The models in Table \ref{tab:models1} are associated with the multilevel binary LR in (\ref{eqn:mlm_lr}) for analyzing binary responses, while the models in Table \ref{tab:models2} are related to the multilevel multinominal LR in (\ref{eqn:mlm_poly}) for analyzing polytomous responses. The models in Table \ref{tab:models1} were fitted via \texttt{lme4} package \citep{bates_linear_2022-1} and the models in Table \ref{tab:models2} were fitted via \texttt{brms} and \texttt{rstan} packages, with R4.4.0 in RStudio. Appendix C.2 documents the R code for this modeling. 

In both the tables, the first three models are for multilevel model building, named \enquote{Model 0,} \enquote{Model 1,} and \enquote{Model 2,} and the last model is the single-level model, labeled with \enquote{Model 3.} Model 0 does not contain explanatory variables in the level-1 submodel, commonly known as \textit{null model} or \textit{empty model} that sets a baseline for understanding the proportion of variance explained by the hierarchical structure of the data \citep{snijders_multilevel_2011}. Model 1 is the exact model used in (\ref{eqn:mlm_lr}) or (\ref{eqn:mlm_poly}), incorporating the latent trait and group membership variables. Model 2 adds the interaction term between the latent trait and group membership, which is the interaction term we drop from the traditional LR, to check how the model performance would change with this interaction. 

\begin{table}
\centering
\caption{Multilevel binary LR: Analyzing binary responses.}
\label{tab:models1}
\begin{tabular}{>{\raggedright\arraybackslash}p{0.15\linewidth} >{\raggedleft\arraybackslash}p{0.15\linewidth} >{\raggedleft\arraybackslash}p{0.15\linewidth} >{\raggedleft\arraybackslash}p{0.15\linewidth} >{\raggedleft\arraybackslash}p{0.15\linewidth}}  
    \toprule
    & \text{Model 0} & \text{Model 1} & \text{Model 2} & \text{Model 3} \\ 
    \midrule
    
    \multicolumn{2}{l}{\textbf{Fixed effects}} & & & \\
    \hspace{3pt} (Intercept) & -1.02$^{***}$ & -2.37$^{***}$ & -2.39$^{***}$ & -1.90$^{***}$ \\ 
    & (0.15) & (0.15) & (0.15) & (0.03) \\ 
    \hspace{3pt} $\theta_p$ &               & 0.25$^{***}$ & 0.25$^{***}$ & 0.25$^{***}$ \\ 
                              &               & (0.00) & (0.00) & (0.00) \\ 
    \hspace{3pt} $g_{p.1}$ &               & -0.12$^{***}$ & -0.09$^{*}$ & -0.07 \\ 
                              &               & (0.02) & (0.04) & (0.04) \\
    \hspace{3pt} $g_{p.2}$ &               & 0.01 & -0.05 & -0.12 \\ 
                              &               & (0.06) & (0.09) & (0.09) \\

    \hspace{3pt} $\theta_p g_{p.1}$ &               &        & -0.01 & -0.00  \\ 
                              &               &        & (0.01) & (0.01) \\ 
    \hspace{3pt} $\theta_p g_{p.2}$ &               &        & 0.01 & 0.01  \\ 
                              &               &        & (0.01) & (0.01) \\ 
    
    \multicolumn{2}{l}{\textbf{Level-2 Variances}} & & & \\
    \hspace{3pt} $\tau_0^2$ & 0.25 & 0.22 & 0.22 &  \\ 
    \cmidrule(lr){1-5}
    
    \hspace{3pt} N & 45827 & 45827 & 45827 & 45827 \\ 
    \hspace{3pt} ICC & 0.07 & 0.06 & 0.06 &  \\
    \hspace{3pt} AIC & 59233 & 46286 & 46288 & 46809 \\ 
    \hspace{3pt} $R^2$ & 0.07 & 0.38 & 0.38 &  \\ 

    \bottomrule
    \end{tabular}
        \begin{tablenotes}  
            \item \hspace{-15pt} \textit{Note}. $^{*}p<0.05$; $^{**}p<0.01$; $^{***}p<0.001$. $g_{p.1}$ and $g_{p.2}$ correspond to $g_p$ when the group membership is 1 and 2, respectively.
        \end{tablenotes}
\end{table}

\begin{table}
\centering
\caption{Multilevel multinominal LR: Analyzing polytomous responses.}
\label{tab:models2}
\begin{tabular}{>{\raggedright\arraybackslash}p{0.15\linewidth} >{\raggedleft\arraybackslash}p{0.1\linewidth} >{\raggedleft\arraybackslash}p{0.1\linewidth} >{\raggedleft\arraybackslash}p{0.1\linewidth} >{\raggedleft\arraybackslash}p{0.1\linewidth} >{\raggedleft\arraybackslash}p{0.1\linewidth} >{\raggedleft\arraybackslash}p{0.1\linewidth} >{\raggedleft\arraybackslash}p{0.1\linewidth} >{\raggedleft\arraybackslash}p{0.1\linewidth}}  
    \toprule
    \multicolumn{1}{c}{ } & \multicolumn{2}{c}{Model 0} & \multicolumn{2}{c}{Model 1} & \multicolumn{2}{c}{Model 2} & \multicolumn{2}{c}{Model 3} \\
    \cmidrule(lr){2-3} \cmidrule(lr){4-5} \cmidrule(lr){6-7} \cmidrule(lr){8-9}
 &  \multicolumn{1}{c}{$Eq_1$}  &  \multicolumn{1}{c}{$Eq_2$}  & \multicolumn{1}{c}{$Eq_1$}  &  \multicolumn{1}{c}{$Eq_2$}  & \multicolumn{1}{c}{$Eq_1$}  &  \multicolumn{1}{c}{$Eq_2$}  & \multicolumn{1}{c}{$Eq_1$}  &  \multicolumn{1}{c}{$Eq_2$} \\

    \midrule
    
    \multicolumn{2}{l}{\textbf{Fixed effects}} & & & \\
\hspace{3pt} (Intercept) & -1.02$^{***}$ & -3.12$^{***}$ & -2.37$^{***}$ & -3.66$^{***}$ & -2.37$^{***}$ & -3.69$^{***}$ & -1.89$^{***}$ & -3.73$^{***}$\\
 & (0.19) & (0.07) & (0.18) & (0.09) & (0.18) & (0.09) & (0.02) & (0.06)\\

\hspace{3pt} $\theta_p$ &  &  & 0.25$^{***}$ & 0.13$^{***}$ & 0.25$^{***}$ & 0.14$^{***}$ & 0.24$^{***}$ & 0.14$^{***}$\\
 &  &  & (0.00) & (0.01) & (0.00) & (0.01) &(0.00) & (0.01)\\

\hspace{3pt} $g_{p.1}$ &  &  & -0.12$^{***}$ & -0.02 & -0.09$^{*}$ & 0.03 & -0.07 & 0.01\\
 &  &  & (0.02) & (0.06) & (0.04) & (0.09) & (0.04) & (0.09)\\

\hspace{3pt} $g_{p.2}$ &  &  & 0.01 & -0.14 & -0.06 & -0.17 & -0.23$^{*}$ & -0.29\\
 &  &  & (0.05) & (0.15) & (0.09) & (0.22) & (0.09) & (0.22)\\

\hspace{3pt} $\theta_p g_{p.1}$ &  &  &  &  & -0.01 & -0.01 & 0.00 & -0.01\\
 &  &  &  &  & (0.01) & (0.01) & (0.01) & (0.01)\\

\hspace{3pt} $\theta_p g_{p.2}$ &  &  &  &  & 0.01 & 0.01 & 0.04$^{*}$ & 0.06\\
 &  &  &  &  & (0.01) & (0.03) & (0.01) & (0.03)\\
    
    \multicolumn{2}{l}{\textbf{Level-2 Variances}} & & & \\
    \hspace{3pt} $\tau_0^2$ & 0.37 & 0.02 & 0.31 & 0.02 & 0.32 & 0.02 \\ 
    \cmidrule(lr){1-9}
    
    \hspace{3pt} N & 47094 & 47094 & 47094 & 47094 & 47094 & 47094 & 47094 & 47094\\ 
    \hspace{3pt} ICC & 0.10 & 0.01 & 0.09 & 0.01 & 0.09 & 0.01 \\  
    \hspace{3pt} WAIC & 70824 & 70824 & 57909 & 57909 & 57915 & 57915 & 58506 & 58506\\
    \hspace{3pt} $R^2$ & 0.00 & 0.00 & 0.23 & 0.23 & 0.23 & 0.23 & 0.23 & \\ 
    \bottomrule
    \end{tabular}
        \begin{tablenotes}  
            \item \hspace{-15pt} \textit{Note}. $^{*}p<0.05$; $^{**}p<0.01$; $^{***}p<0.001$. $g_{p.1}$ and $g_{p.2}$ correspond to $g_p$ when the focal group's group membership value is 1 and 2, respectively. $Eq_1$ and $Eq_2$ refer to \enquote{Equation 1} and \enquote{Equation 2} when the targeted response is 1 and 2, respectively. They share the same WAIC and $R^2$ values for each model.
        \end{tablenotes}
\end{table}

The modeling results reveal several key insights regarding the model performance. Model 0 serves as the baseline, showing the proportion of variance attributable to the hierarchical structure of the data. Its intraclass correlation coefficient \citep[ICC;][]{raudenbush_hierarchical_2002} 
is 0.07 in Table \ref{tab:models1} and 0.10 in Table \ref{tab:models2}, indicating that 7\% of the response variability stemmed from differences between education levels for the first data set and this proportion is 10\% for the second data set. These values highlight that accounting for the hierarchical structure in DIF analyses matters with both the data sets. 

Model 1 displays a significant negative effect of being from high-density areas on the probability of responding \enquote{Yes} to the item, for both the data sets. This suggests that the selected survey item has population-density-associated DIF, as individuals residing in high-density areas were less likely to provide the targeted response to the selected item compared to those in low-density areas, after accounting for all these individuals' latent trait about depression. Model 3, the single-level model, does not reveal a significant effect of high population density, which leads to the conclusion that there is no DIF when comparing high- and low-density areas, contrasting with the result from Model 1. This difference underscores the importance of model selection when conducting DIF analyses.

Overall, Model 1 outperforms others, based on model performance statistics. Its intercept variance $\tau_0^2$ is among the lowest in both the tables, indicating an advantage of Model 1 in capturing more data variability. Additionally, Model 1 has the lowest value in the model fit index, namely the Akaike information criterion (AIC) in Table \ref{tab:models1} and the widely applicable information criterion (WAIC) in Table \ref{tab:models2}. In addition, Model 1 has the highest conditional $R^2$ \citep{nakagawaGeneralSimpleMethod2013b}, signifying that the entire model, with random effects and fixed effects in combination, explains 38\% of the variability in the first data set and 23\% in the second data set.

\section{Discussion}

Multilevel modeling we describe in this article is a useful tool for examining DIF in the data from health-related constructs or measures, such as self-reported health survey data. By leveraging this approach, researchers can gain a more accurate understanding of DIF and glean valuable insights into health survey data, ultimately enhancing the precision and reliability of health research findings. Built on prior research that highlights the benefits of multilevel modeling in DIF analysis, our study provides a practical application of how this approach can be used with self-reported health survey data. 

Our empirical analyses cover the DIF analysis of a dichotomous item and a polytomous item, using multilevel binary LR and multlevel multinominal LR in addition to their single-level counterparts. Rather than making a blanket recommendation in favor of multilevel models, we intend to raise awareness of how to use multilevel modeling to analyze DIF and how to choose the appropriate model when a series of competing models are available. Different from some other studies that evaluated models via Monte Carlo simulation \citep[e.g.,][]{svetina_valdivia_detecting_2024}, our empirical analysis examines a series of competing models based on model performance statistics. Our findings indicate that multilevel modeling more effectively captures variability in our data compared to single-level modeling. Future research could explore the generalizability of our findings by applying multilevel DIF analysis to other self-reported health survey data. Also, researchers might consider comparing our models with other models capable of multilevel modeling for DIF analysis with polytomous items or more than two groups, ideally models without IRT-related assumptions such as ordinal LR \citep{crane_rapid_2007,zumbo_handbook_nodate} or nonparametric measures \citep[e.g.,][]{zwick_evaluating_1996}. 

More research is needed to support multilevel modeling. First, literature limitations exist regarding model construction, ICC calculation, and generalizability \citep{chenModelingItemBias2023a}, despite existing efforts to explore multilevel DIF analysis \citep[e.g.,][]{french_hierarchical_2010,french_transforming_2015,french_multilevel_2019}. Second, it would be worthwhile to examine the cost-benefit balance of multilevel modeling for DIF analysis in dynamic health survey settings, such as using Monte Carlo simulations to systematically evaluate model performance under a wide range of data structures. These simulations can vary factors such as ICC, number and size of clusters, the magnitude and proportion of DIF items, and data distribution within each cluster \citep{snijders_multilevel_2011}.
Studies by \citet{moineddin_simulation_2007} and \citep{woods_langer-improved_2013} provide valuable groundwork, but further exploration is necessary. Third, scrutinizing model performance under various realistic scenarios that mimic real-world health survey data is essential. This includes exploring the impact of factors like missing data patterns (missing completely at random, missing at random, or missing not at random) and different types of DIF (uniform vs. non-uniform) on model performance. By addressing these limitations, we can solidify the role of multilevel modeling in DIF analysis for health research, ultimately leading to more robust and reliable estimates captured through self-reported surveys.

\backmatter

\section*{Declarations}

\bmhead{Funding}
The authors declare that no funds, grants, or other support were received during the preparation of this manuscript.

\bmhead{Competing Interests}
The authors have no relevant financial or non-financial interests to disclose.

\bmhead{Availability of Data and Materials}
The NSDUH data and the codebook used in this article can be downloaded here: https://www.datafiles.samhsa.gov/dataset/national-survey-drug-use-and-health-2022-nsduh-2022-ds0001

\bibliography{bibliography}

\end{document}